\documentclass{aa7}  

\usepackage{graphicx,amssymb}
\usepackage[varg]{txfonts}
\begin{document}
   \title{Towards equation of state of dark  energy from quasar monitoring: Reverberation strategy}

   \author{B. Czerny\inst{1}  \and K. Hryniewicz\inst{1} \and I. Maity\inst{1} \and A. Schwarzenberg-Czerny\inst{1,2}  \and P.T. \. Zycki\inst{1}  \and M. Bilicki\inst{3}
          }
   \institute{Nicolaus Copernicus Astronomical Center, Bartycka 18, 00-716 Warsaw, 
                          Poland 
\and
Adam Mickiewicz University Observatory, ul. Sloneczna 36, 60-286, Poznan, Poland
\and
Astrophysics, Cosmology and Gravity Centre,
Department of Astronomy,
University of Cape Town,
Private Bag X3,
Rondebosch 7701,
Republic of South Africa
             }

   \date{Received ............; accepted ..............}

 
  \abstract
   {High redshift quasars can be used to constrain the equation of state of dark energy, as a complementary tool to Supernovae type Ia, especially at $z>1$.}
   {The method is based on determination of the size of the Broad Line Region from the emission line delay, 
determination of the absolute monochromatic luminosity either from the observed statistical relation or 
from a model of 
the formation of the Broad Line Region, and determination of the observed monochromatic flux from photometry. 
This allows to obtain the luminosity distance to a quasar independently from its redshift. The accuracy of 
the measurements is however, a key issue.}
   {We model the expected accuracy of the measurements by creating artificial quasar monochromatic 
lightcurves and responses from the Broad Line Region under various assumptions about the variability of a quasar, 
Broad Line Region extension, distribution of the measurements in time, accuracy of the measurements and the 
intrinsic line variability.}
   {We show that the five-year monitoring of a single quasar based on Mg II line should give an accuracy of 0.06 - 0.32 magnitude 
in the distance modulus which will allow to put new constraints on the expansion rate of the Universe at high redshifts. Successful monitoring of 
higher redshift quasars based on CIV lines requires proper selection of the objects to avoid sources with much higher level of the intrinsic variability of CIV in comparison with Mg II.}
   {}

\authorrunning{Czerny et al.}
\titlerunning{Reverberation strategy for high redshift quasars}

   \maketitle
%

\section{Introduction}
The key issue of the present day cosmology is  to determine the properties of the dark energy that drives the accelerated expansion of the Universe  (for a review, see Frieman, Turner \& Huterer 2008, Amendola \& Tsuikawa 2010, Li et al. 2011, Mattarese et al. 2011).   
The need for the cosmological constant was already posed in early 1980s (e.g. Peebles 1984), but the arguments accumulated slowly. Situation started to change after the release of  the COBE cosmic microwave background (CMB) measurements , particularly when combined with other constraints coming from the large structure formation (e.g. Kofman et al. 1993 estimated that $\Omega_{\Lambda} \sim 0.3 - 0.4$). 
The convincing evidence for $\Lambda > 0$ came with the observations of Supernovae Ia (Riess et al. 1998, Perlmutter et al. 1998) , which lead to the emergence of the standard cosmological model, $\Lambda$CDM.  
There was a considerable quantitative progress since that time, with the emergence of precision cosmology. 
The presence of the dark energy is  now well supported observationally, and most of the results are still consistent with
its simplest interpretation as a cosmological constant. However, a  number of dark energy models has been proposed,
giving more general predictions for the equation of state, $P = w \rho c^2$,  than the simple cosmological constant, which corresponds to $w = -1$. An example of such a more complicated model is the quintessence model, where  $w$ is constant in time and lies in the range $ -1 <w < -1/3$, or the phantom dark energy with $w < -1$. In even more general models $w$ can change 
with the time/redshift (e.g. Chevallier \& Polarski 2001).

 Most recent results for $z \sim 1$ Supernovae of Suzuki et al. (2012) imply $\Omega_{\Lambda} = 0.729 \pm 0.014$ for a $\Lambda$CDM model, and $w = -1.013^{+0.068}_{0.073}$ (68 \% confidence level) for a flat wCDM model. Similar numbers have been obtained based on CMB measurements: WMAP 9-year data analysis gives  $\Omega_{\Lambda} = 0.721\pm0.025$ (Hinshaw et al. 2012), whereas Planck 2013 results point to somewhat lower value ($\Omega_{\Lambda} = 0.686 \pm 0.020$) (Ade et al. 2013,  Table 2). Both experiments show additionally that the simplest LambdaCDM model with w=-1 is a very good fit to the data, although various degeneracies do not allow to exclude other dark energy equations of state.  
Additional constraints on the expansion history of the Universe come from the large-scale structure studies, for instance via Baryon Acoustic Oscillation (Eisenstain et al. 2005; Sanchez, A.G. et al. 2012) or redshift-space distortions (Guzzo et al. 2008). New methods are being constantly suggested, like those based on weak lensing of galaxies (statistical approach: Jain and Taylor 2003, individual sources: Biesiada \& Piorkowska  2009), combined X-ray emission and Sunyaev-Zel'dovich effect in galaxy clusters (see Benson et al. 2013 and the references therein), gamma-ray bursts (e.g. Dainotti et al. 2011) and standard sirens in gravitational waves (suplemented with redshift, Schutz 1986, or without it, Seto et al. 2001).  Combining such methods allows for independent checks of the validity of the standard cosmological model. However, as far as the dark energy equation of state and its redshift evolution are concerned, those cannot be well constrained from low-redshift and CMB data only: one needs to directly probe the expansion rate at $z \gtrsim 1$ to  effectively study the time evolution of w. At these redshifts, however, SNeIa are very difficult to observe, due to both a considerable shift of their spectra towards the infrared and cosmological dimming (the latter $\propto (1+z)^4$).

Watson et al.\ (2011) proposed that Active Galactic Nuclei (AGN) can also be used as  
probes of the dark energy, complementary to the study done with other methods. The basic idea is the following. The
absolute size of the Broad Line Region (BLR) in AGN can be measured from the time delay between 
the variable continuum
emission from the nucleus and the emission lines in the optical/UV band (e.g. Blandford \& McKee 1982; 
Peterson 1993; Sergeev et al.\ 1994; Wandel et al.\ 1999). This reverberation technique based on the
H$_\beta$ line 
was succesfully used for low redshifts objects (up to $z=0.292$; Kaspi et al.\ 2000, Bentz et al.\ 2009ab). 
The size of the BLR is related to the absolute value of the source monochromatic luminosity. 
This was found
observationally (Kaspi et al.\ 2000; Peterson et al.\ 2004; Bentz et al.\ 2009a), but the relation was 
later theoretically 
explained by Czerny \& Hryniewicz (2011) as corresponding to the formation of the inner radius of BLR 
at fixed temperature of 1000 K in all sources, independently from the black hole mass and accretion rate.
This result points toward the important role of dust in the accretion disk atmosphere in formation of BLR,
connecting it with the dust sublimation temperature. Having the absolute value of the monochromatic luminosity,
and measuring the observed monochromatic luminosity, we can determine the luminosity distance to the source.
Summarizing, the
measurement of the observed monochromatic flux and of the time delay between continuum and an BLR emission line
for a number of sources allows to construct the diagram of the 
luminosity distance as a function of redshift, and to test the cosmological models.

The use of an independent tracer of the dark energy is important. Various methods are being used or will be used
to trace the dark energy, but each of them can have certain bias. Supernovae luminosity may be 
subject to systematic cosmological evolution, due to the evolution
of the chemical content of the Universe. Quasars show roughly solar or slightly super-solar metallicity even to redshifts
of a few (e.g.\ Dietrich et al.\ 2003a,b; Simon \& Hamann 2010; De Rosa et al.\ 2011; Gall et al.\ 2011), 
probably due to the correlated evolution of starburst and quasar activity. Therefore, the
relation between the size of the BLR and monochromatic luminosity, reflecting dust properties, should
not contain strong evolutionary trends with the redshift. Other methods, like galaxy clusters, weak 
gravitational lensing and galaxy angular clustering (baryon acoustic oscillations) can also contain some bias since
the clusters are not spherically symmetric and relaxed, and galaxy clustering may be the subject of magnification
bias. If numerous independent  methods are used, the results will be more reliable.

We have recently started an observational campaign with the South African Large Telescope (SALT) 
to collect data for reverberation studies of a sample of 7 objects. We are planning to use the low 
ionization line of Mg II 2798\AA, which is suitable for studies of objects in the redshift range 
from 0.9 to 1.4. So far only a few authors claimed the measurement of the delay in Seyfert galaxies 
using Mg II or CIV line using HST or IUE observations
(Clavel et al. 1991 for NGC 5548, $z = 0.017175$  using Mg II and CIV,  Reichert et al. 1994 for NGC 3783, $z = 0.00973$,  using Mg II and CIV, Wang et al. 1997 and O'Brien  et al. 1998 for 3C 390.3, $z = 0.0561$, using CIV line; Peterson \& Wandel 1999 for NGC 5548, $z = 0.017175$ using CIV, Peterson et al. 2004 for Fairall 9, $z = 0.047016$, using Mg II line, Metzroth et al. 2006 for NGC 4151, $z = 0.003319$, using CIV and Mg II). For higher redshift objects  Chelouche
et al.\ 2012 were able to measure the delay for MACHO 13.6805.324, $z = 1.72$, from photometric method, but in this case the measurement likely
represents the response of the iron pseudo-continuum. Woo (2008) showed that using Mg II
line is {likely an attractive} possibility but the two year monitoring program was of course too short to attempt any
delay measurement. Reverberation studies of even higher redshift 
quasars, based on the CIV 1549\AA~ line,  were recently done with the HET telescope by Kaspi et al.\ (2007), 
but the authors generally were not successful in the measurement of the time delays, with a tentative delay 
measurement of 595 days for one object (S5 0836+71) out of six. 
Therefore, in this paper we perform a detailed analysis of the requirements for a successful monitoring program.


\section{Method}

AGN are highly variable in all energy bands (for a review, see Ulrich et al. 1997), including optical band 
(e.g. Matthews \& Sandage 1963; 
Hawkins 1983; Barbieri et al. 1988; Cristiani et al. 1996; 
Vanden Berk et al. 2004; Bauer et al. 2009; MacLoad et al. 2012). Although best studies were done in the X-ray band, there
were also several attempts to describe quantitatively the character of the optical variability. The power density
spectra (PDS) in the optical band for NGC 4151 were constructed by Terebizh, Terebizh \& Biryukov (1989) and by 
Czerny et al.\ (2003) in the long time-scale, and by Edelson et al.\ (1996) in the short time-scale; for NGC 5548
by Czerny et al. (1999).  



 Therefore our idea here is to create simulated datasets of a fiducial high redshift quasar to 
estimate the role of various parameters of an observational campaign (e.g., number and separations 
of observations, duration of the campaign) and expected quality of data, in the accuracy of the 
determined reverberation time delay. This should help in developing
the optimum strategy of such reverberation studies campaign.

We use the  Timmer \& K\" onig (1995) algorithm for generating an artificial continuum lightcurve of a 
quasar based on assumed shape of the power density spectrum (PDS).
We then exponentiate the resulting lightcurve, following Uttley et al. (2005), since this reproduces the 
log-normal distribution characteristic for high energy lightcurves of accreting sources.
We assume that the PDS has a shape of a doubly-broken power law, although the long term variability is highly 
uncertain, but this allows us to test whether there are any effects of the power leak toward shorter frequencies. 
We fix the longest frequency slope at zero, while the two remaining slopes, $s_1$ and $s_2$, and the break 
frequencies,  $f_1$ and $f_2$, are in general the free parameters of the simulations. 
We expect that they are mostly affected by the value of the black 
hole mass of the monitored quasar, since in the X-ray band the black hole mass is the key parameter of the PDS 
(Hayashida et al.\ 1998; Czerny et al.\ 2001; Papadakis 2004, Zhou et al.\ 2010; Ponti et al.\ 2012), 
although the effect of the Eddington ratio and/or spectral state is an issue (e.g., McHardy et al.\ 2006; 
Nikolajuk et al.\ 2009). 
We fix these parameters at: $s_1=1.2$, $s_2=2.5$, $f_1=0.012$yr$^{-1}$ 
and $f_2=0.18$yr$^{-1}$. These values were estimated from the optical PDS of NGC 5548 (the intermediate slope, $s_1$ 
and the break frequencies, $f_1$ and $f_2$), while the high frequency slope, $s_2$ was taken from the {\it Kepler} 
quasars monitoring campaign, after correcting the slope for the subtraction of the linear trend 
(Mushotzky et al. 2011). The break frequencies were then scaled to a higher expected black 
hole mass, and for the redshift. The overall dimensionless amplitude of the PDF is also a model parameter, 
but again in most simulations it was fixed at 0.1 and 0.3, which correspond to the dimensionless dispersion 
in the infinite photometric lightcurve (i.e., of duration above 1 000 yr) without any measurement error  of 
0.32  and 1.61, correspondingly.
The created lightcurves have the total standard length of 30 years and are densly spaced (1 day).

Next we model the response of the BLR (meant to be represented by the Mg II 2798\AA ~line) 
to the variability of the continuum.
In the simplest approach the line lightcurve is created by convolving the continuum lightcurve with a Gaussian kernel. 
The center of the Gaussian represents the line delay after the continuum, which is a free 
parameter of the model, and is of the order of a few hundred days.
The Gaussian width models the smearing of the response due to the finite extension of the BLR. 
Most of the reprocessing is done by the BLR part located on the far side of the black hole 
since the clouds on the side to the observer are dark and so they 
do not emit efficiently. Therefore, in most computations 
we assume the Gaussian width of the order of 10 percent of the time delay, since such parameters 
recover well the measured accuracy of the time delay for NGC 5548. However, we also experiment with other values of the Gaussian width as well as with more complex BLR response.

Obviously, the photometric and spectroscopic monitoring with SALT telescope cannot be done with 1 day spacing, 
so in the next step we select observational points from the two light curves, assuming a realistic 
observational pattern and simulating also the measurement errors. In most cases we assume that 
the object cannot be observed for about 3 months every year (it is then out of the observational 
window of SALT), and the gap is the same for photometric and spectroscopic lightcurve. 
For photometric monitoring we assume that the object will be observed roughly every two weeks, 
with random dispersion of 3 days. For spectroscopic measurements, we assume 5 SALT observations every 
year, with random dispersion of 14 days. We can of course vary these parameters as well.

Finally, we have to account for possible uncertainties of both the spectroscopic and photometric measurements.
Therefore, the value of the lightcurve at each observational point is replaced with a random value obtained 
from a Gaussian distribution with the mean equal to this value and the dispersion corresponding to the 
assumed 1 sigma error. The errors for the line measurement are generally assumed to be larger 
than the errors of the photometric measurements. As a default, we use the conservative estimate of 5\% accuracy
in photometry and 10 \% accuracy in spectroscopy.

Having the two curves, we now measure the time delay of the spectroscopic lightcurve with respect to the 
photometric one, for example using the interpolated cross-correlation function (ICCF) as usually done 
for the real data (e.g. Gaskell \& Peterson 1987). 
We interpolate the spectroscopic measurements to the dates of the photometric measurements, and we use 
the maximum of the ICCM to determine the delay. In order to determine the expected accuracy of the measurement, 
for each set of the all other parameters we analyze 1000 random realizations of the process, and we create 
the histogram of the measured delays. The comparison of this histogram with the assumed delay characterizes 
the expected quality of the observational campaign. 
In particular, we use the median of the histogram to give the best value of the delay, 
and for the dispersion we give the dispersion around the known expected value. In addition we give 
the spread, defined as a half range between the $1 - 0.158$ and 0.158 quantiles of the histogram. 
For the Gaussian distribution it would correspond to $1 \sigma$ error. We also experiment with alternative 
methods to measure time delay.


The whole procedure should mimic well the actual reverberation measurements performed for numerous less 
distant sources (e.g., Kaspi et al.\ 2000; Peterson et al.\ 2002; Bentz et al.\ 2009ab; Denney et al.\ 2009; 
Walsh et al.\ 2009; Denney et al.\ 2010; Barth et al.\ 2011; Stalin et al.\ 2011; 
Dietrich et al.\ 2012; Grier et al.\ 2012a, Grier et al.\ 2012b).  

We tested the method using the past monitoring of NGC 5548. The optical power spectrum for this source was described as a power law with two breaks, with the slopes as above but lower frequency breaks ($f_1=0.11$yr$^{-1}$ 
and $f_2=3.6$yr$^{-1}$). We assumed 0.04 for the dimensionless level of variability. For the observational setup, we considered sparse monitoring described by Peterson et al. (2002) and the dense short Lick monitoring presented by Walsh et al. (2009) and Bentz et al. (2009b). In the case of the first monitoring, we used setup for Year 8 (1996), with the measured delay $15.3$ day. We took  5\% and 4 \% for photometry and spectroscopy error, correspondingly. The computations returned the delay measurement, $15.0 \pm 1.9$ days, with the rms of 8.5 \% and 10.2 \% for photometry and spectroscopy, close to observed values (15 \% and 11 \%, correspondingly). 
The second monitoring lasted for 68 days, with 51 line measurements, we assumed the measurements errors of 2\% and 6 \% for photometry and spectroscopy, representative for this monitoring, and we assumed the same overall variability normalization as before, but much shorter delay (4.2 days; the source was then in a very low state). We recover the delay $4.2 \pm 1.8$ days, with $F_{var}$ in the line and in the continuum of 8.5\% 5.4\%, again similar to the observed values.

With our Monte Carlo approach we are then able to study which elements of the campaign are the most 
important ones and where there are any possibilities for an improvement. These we hope will be of 
significant help in planning a c.a.\ 5 years long observational project. 


\section{Results}


We model the expected accuracy of the measurement of the emission line delay with respect to 
the continuum for a single object (high redshift quasar) monitoring. The problem is important 
because the variability of AGN are of a red noise type, with a broad range of timescales and 
a strong contributions from the timescales much longer than the observed lightcurves. 
The Monte Carlo approach is effective in assessing the likely results as well as possible caveats 
since in the real monitoring we have to work just with one realization of the process.

The basic model (hereafter model A) is adjusted to our already started observational campaign. 
The feasibility study has been done for a quasar  Q 2113-4538 ($z = 0.946$). The high redshift of the source 
ensures that the Mg II 2798\AA~ line is within the spectroscopy frame.  The quasar comes 
from the Large Bright Quasar Sample (LBQS; e.g., Hewett et al.\ 1995), the black hole mass in this object 
was estimated to be  $1.5 \times 10^9 M_{\odot}$ (Vestergaardt \& Osmer 2009), and the expected time 
delay of the Mg II line (from the Kong et al. 2006 formula) is 660 days, if the additional increase due to the 
redshift is included. The spectroscopy is being done with the SALT and the source is visible for about 
9 months. We plan to perform 5 observations per year and three observations were already performed. 
The photometry will be performed more frequently, 
with smaller telescopes, approximately every 2 weeks. For the time being, the collaboration with the OGLE 
team has been secured.

We have already analyzed the first three spectra for Q 2113-4538 obtained with the SALT telescope, one in Dark Moon 
conditions, one in the Grey Moon, and one in the Bright Moon conditions, and the statistical error in
 the Mg II 2798\AA ~
line equivalent 
width (EW) of the two spectra were 0.3\%. However, possible uncertainty in determination of the level 
of the Fe II pseudo-continuum can contribute to the estimate of the Mg II line EW, and some other 
systematic errors are also likely to be present. 
Therefore in the model A we  adopt 10\% as the actual measurement 
accuracy of the line intensity in the basic model. The photometric accuracy of continuum measurement 
in the basic model is set to 5\%. 
We give all the parameter values in Tables~\ref{table:setup2} and ~\ref{table:setup1}. 

Model A represents our reference setup of the campaign, while other models (from B to E)
illustrate the dependence of the final results on the details of the campaign. 
We repeat the computations for two values of the amplitude of the continuum variability: 
a most likely one (first case) and a larger amplitude (second case).

First we test the ICCF (maximum value) method to determine the time delay in the simulated data, since this method is 
most often used in real data analysis. 

Running 1000 realizations for model A we properly recover the assumed time delay of 660 days, 
but the median value has a noticeable error: $605 \pm 285$ days for low variability case and $605 \pm 268$ 
for the other case. Such a measurement is not quite satisfactory, although if the monitoring 
is done for a number of objects, the statistical error in determination of the dark energy 
distribution will be reduced. However, there are also possibilities of reducing the error of 
the delay measurement for a single object with a more careful approach to the monitoring.

First, we test the importance of the accuracy of measurements. Model B represents the situation 
when the individual measurement errors are significantly reduced, with the remaining organization of 
the campaign schedule unchanged. This, however, does not reduce significantly the uncertainty of 
the final result. Denser photometric or spectroscopic monitoring does not reduce the dispersion 
around the delay, either (see Table~\ref{table:setup1}), if the other parameters remain unchanged. 
Only extending the monitoring to much longer timescales, of order of 20 years, results in significant 
improvement. This is due to the strong red noise type variability of the source.
For the assumed parameters, model A shows the dispersion in the photometric points of 0.16 and 0.57, 
and the dispersion in the spectroscopic points of 0.18 and 0.52 for the low and high variability level,
respectively. Still, long timescale trends dominate the source variability
and this leads to a significant fraction of the lightcurves with time delay measured very badly. 
We analyze this problem in detail below.

%
\begin{table*}
\caption{Parameters of the exemplary models and the accuracy of the delay recovery for variability amplitude 0.1, assumed delay 660 days}   
\label{table:setup2}      
\centering                          
\begin{tabular}{l r r r r r r r r}        
\hline\hline      
Model                                &    A & B & C & D & E  \\
Parameter               &    \\    
\hline                        
Photometric~ accuracy                & 5 \%     & 1\%      & 5 \%     & 5 \%     & 5 \%     & \\   
Spectroscopic~ accuracy              & 10 \%    & 1\%      & 10 \%    & 10 \%    & 10 \%    & \\
Width of the BLR response            & 10 \%    & 10\%     & 10 \%    & 10 \%    & 10 \%    & \\
No of photometric points/year        & 18       & 18       & 18       & 100      & 18       & \\
No of spectroscopic points/year      & 5        & 5        & 5        & 5        & 18       & \\
Dispersion of the photometry date    & 3 days   & 3 days   & 3 days   & 1 day    & 3 days   & \\
Dispersion of the spectroscopy date  & 14 days  & 14 days  & 14 days  & 14 days  & 3 days   & \\
Observational gap/year               & 3 months & 3 months & 3 months & 3 months & 3 months & \\
Campaign duration                   & 5 years  & 5 years  & 20 years & 5 years  & 5 years  & \\
\hline                        
ICCF \\
Obtained delay                        & $605\pm285$ & $605 \pm 251$  & $660 \pm 94$   & $605 \pm 279$  &  $605 \pm 246$ &\\
Spread                                & $220$       & $192$          & $55$           & $220$          &  $165$         &\\
\hline                        
$\chi^2$ \\
Obtained delay                        & $671\pm113$ & $652 \pm 20$  & $673 \pm 31$  & $671 \pm 111$  &  $671 \pm 76$ &\\
Spread                                & $84$        & $9$           & $40$          & $93$           &  $47$         &\\
\hline                        
ZDCF \\
Obtained delay                        & $622\pm200$ & $666 \pm 38$  & $661 \pm 405$  & $604 \pm 200$  &  $641 \pm 206$ &\\
Spread                                & $141$       & $20$          & $62$           & $137$          &  $119$         &\\
\hline                                   
\end{tabular}
\end{table*}
%
 
%
\begin{table*}
\caption{Parameters of the exemplary models and the accuracy of the delay recovery for variability amplitude 0.3, assumed delay 660 days}   
\label{table:setup1}     
\centering                          
\begin{tabular}{l r r r r r r r r}        
\hline\hline      
Model                                &    A & B & C & D & E \\
Parameter               &    \\    
\hline                        
Photometric~ accuracy                & 5 \%     & 1\%      & 5 \%     & 5 \%     & 5 \%     & \\   
Spectroscopic~ accuracy              & 10 \%    & 1\%      & 10 \%    & 10 \%    & 10 \%    & \\
Width of the BLR response            & 10 \%    & 10\%     & 10 \%    & 10 \%    & 10 \%    & \\
No of photometric points/year        & 18       & 18       & 18       & 100      & 18       & \\
No of spectroscopic points/year      & 5        & 5        & 5        & 5        & 18       & \\
Dispersion of the photometry date    & 3 days   & 3 days   & 3 days   & 1 day    & 3 days   & \\
Dispersion of the spectroscopy date  & 14 days  & 14 days  & 14 days  & 14 days  & 3 days   & \\
Observational gap/year               & 3 months & 3 months & 3 months & 3 months & 3 months & \\
Campaign duration                   & 5 years  & 5 years  & 20 years & 5 years  & 5 years  & \\
\hline                        
ICCF \\
Obtained delay                        & $605\pm268$ & $605 \pm 260$  & $660 \pm 126$  & $605 \pm 261$  &  $660 \pm 247$ &\\
Spread                                & $220$       & $220$          & $28$           & $220$          &  $165$         &\\
\hline                        
$\chi^2$ \\
Obtained delay                        & $671\pm51$ & $652 \pm 17$  & $673 \pm 14$  & $652 \pm 48$  &  $652 \pm 29$ &\\
Spread                                & $28$        & $9$            & $13$           & $28$           &  $28$          &\\
\hline                        
ZDCF \\
Obtained delay                        & $662\pm106$ & $666 \pm 39$  & $662 \pm 30$  & $660 \pm 104$  &  $667 \pm 112$ &\\
Spread                                & $41$        & $18$          & $26$          & $38$           &  $38$          &\\

\hline                                   
\end{tabular}
\end{table*}
%

   \begin{figure}
   \centering
   \includegraphics[width=0.78\hsize]{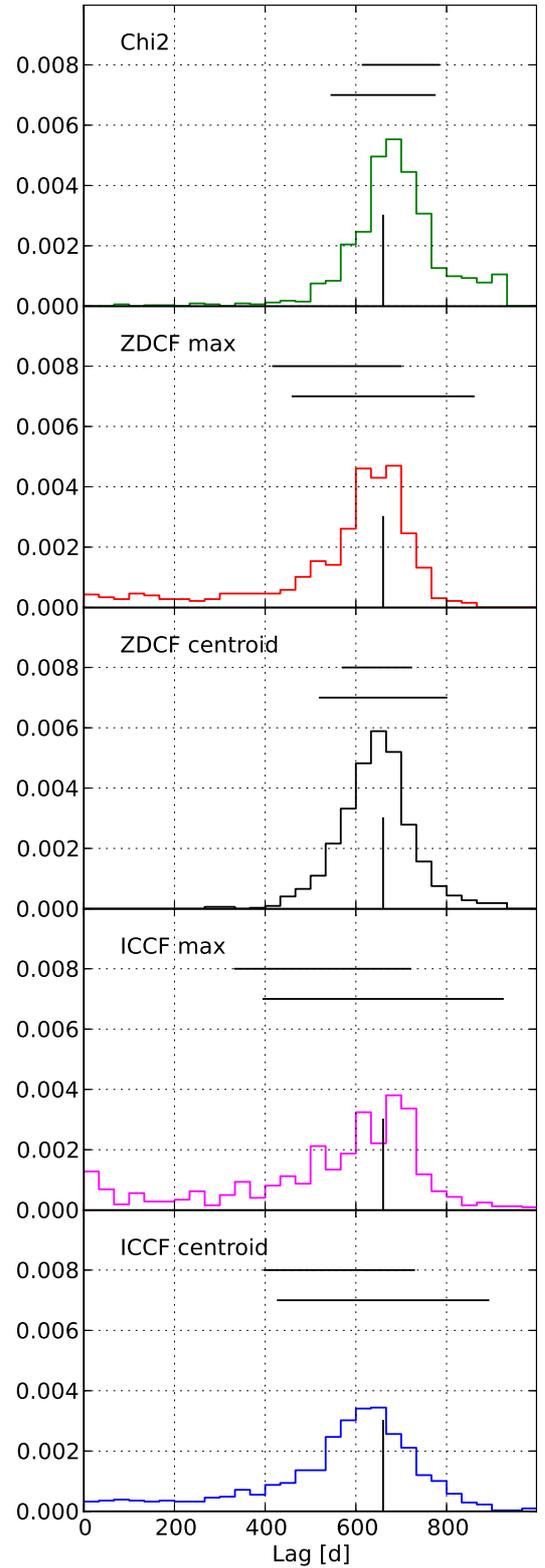}
   \caption{The normalized histogram of the time delays in 1000 realizations of the model A, low variability, for various methods: $\chi^2$ (upper panel), ZDCF and  ICCF method (lower panel). For  ZDCF and ICCF we show results both for maximum and Gaussian centroid method of delay determination, for other methods we give the maximum method. 
{\bf The vertical bar marks the assumed lag value.
The bottom horizontal bar marks standard deviations of the recovered values from the assumed one.
The top horizontal bar marks 0.25-0.75 quantile spread of the recovered histogram}.
Although the average value 
describes well the true delay inserted into the model, there is a considerable tail of much shorter or much longer values which contribute considerably 
to the overall dispersion.}
              \label{fig:histo_ICCF}%
    \end{figure}
%
%

%
   \begin{figure}
   \centering
  \includegraphics[width=1.15\hsize]{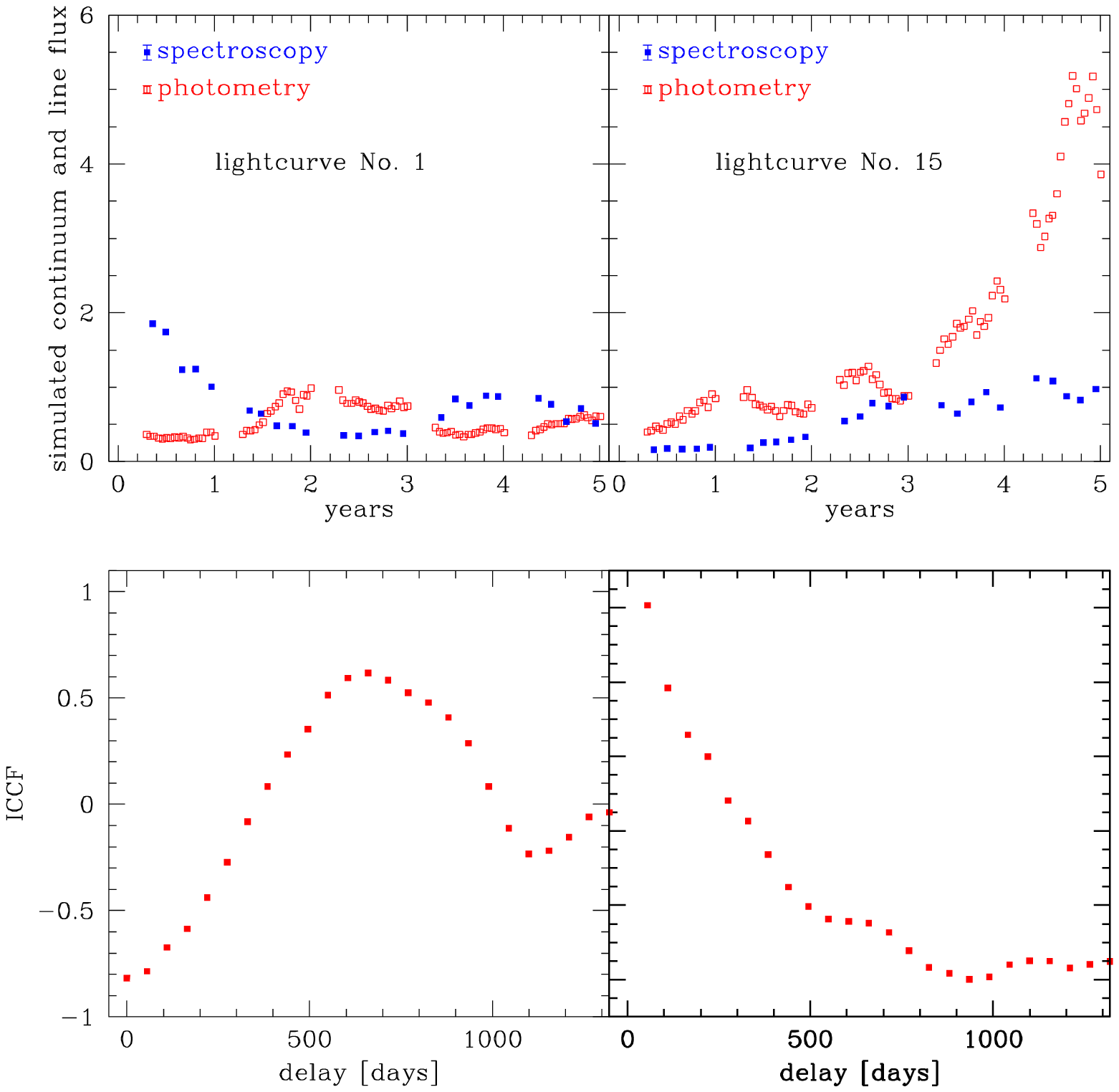}
   \caption{Two examples of the artificial photometric {\bf (filled blue squares)} and spectroscopic {\bf (open red circles)}lightcurves and the corresponding plots of the ICCF, for model A. {\bf Errorbars in the lightcurves (examples shown in the corners) are not much larger than the points.}}
              \label{fig:ICCF}%
    \end{figure}
%
%

\subsection{Dependence on the delay measurement method}

\subsubsection{ICCF problem}

The distribution of the delays obtained for 1000 realizations of the model A, 
low variability level is shown in Fig.~\ref{fig:histo_ICCF}. 
The distribution contains significant tail of measured time delays which are shorter 
than the assumed value. The presence of the tail skews the distribution 
(the mean delay is 549 days, while the median is 605), which enhances the expected 
error of the time delay measurement. 
If we simply neglect the solutions with the delay smaller than 500, then the 
resulting delay is $692 \pm 165$; if in addition we skip the delays longer than 
820 days, the estimated delay is $626 \pm 83$, i.e.\ the error is much lower 
than previously, but we then skip 32.8\% of the cases.
For the higher variability level the longer delay tail is less important, but
the shorter delay tail is very important.  If we use the Gaussian centroid instead of the maximum,
the histogram of the delays is more smooth and the dispersion somewhat lower, but in several cases
we do not obtain any solution to the delay measurement.

Thus with ICCF method, the tails of the distribution strongly affect the results.  
In the real situation we will have just one lightcurve for a single object, 
so there is a considerable probability that the measurement for a single lightcurve 
will be completely wrong. We therefore should characterize the probability of the proper delay 
recovery much better. For a simple Gaussian distribution, one parameter describes
both spread of the distribution core and extent of its tails. The former relates
to precision of estimates, the latter describes reliability i.e.\ probability of
a gross error. Delay statistics\footnote{here statistics means a function of
data, used for estimates} 
based on our simulations have strongly non-gaussian
distribution, hence at least two parameters are needed to characterize their
precision and reliability. For the latter characteristics we
selected standard deviation (s.d.), for the former half spread of symmetric
quantiles limiting central 68.3\% of the distribution. 

In order to illustrate better the source of such large errors we plot two examples of 
the lightcurve realization for model A, and the corresponding ICCF plots (see Fig.~\ref{fig:ICCF}).
The problem is even more severe if the overall variability amplitude is assumed to be lower 
although in this case the large outbursts 
happen less frequently. The ICCF still frequently returns spurious results when the lightcurves 
are dominated by a single monotonic event (flare) 
in the observed window. In addition, lower amplitude makes the observational errors relatively 
more important, so the overall accuracy of the delay determination is lower. 

This suggests a considerable space, and actual need, for improvement. Therefore, 
we consider various alternatives to the ICCF method.
To evaluate performance of the delay statistics we employ two criteria:
its bias and the error. Inspection of histograms obtained from our
simulation reveals asymmetric distribution of the values of the
statistics. Hence we prefer median rather than
average value for estimation of bias. For an arbitrary distribution the former yields equal
probability of underestimated and overestimated result. For estimation
of error we employ standard deviation with respect to the assumed delay
value of 660 days.
As it depends on quadratic norm of residuals it is rather sensitive to
gross errors. As we aim to minimize the gross errors, the standard
deviation seems the right choice.
However, to better characterize our resulting distribution we additionally
list the half spread of their 1-0.158 and 0.158 quantiles. For the
Gaussian distribution it would correspond to half of the $\pm 1\sigma$
range, i.e. to $1\sigma$. A value of the spread small compared to the
standard deviation indicates a distribution with the narrow core and
broad wings. Note that generally for an asymmetric distribution center
of the spread range does not coincide with the median or average.
The corresponding values for all the setups 
are given in Table~\ref{table:setup2} and Table~\ref{table:setup1}.

\subsubsection{$\chi^2$ method}

In this method we shift the spectroscopic curve and interpolate to 
the photometric time measurements as before, but, instead of computing ICCF,
 we minimize the function
\begin{equation}
\chi^2 = {1 \over N} \sum\limits_{i=1}^N {{(x_i - A y_i)^2} \over {\delta x_i^2 + A^2 \delta y_i^2 }},
\end{equation}
where $x_i$ is the photometric flux, $y_i$ is the interpolated line flux, 
$\delta x_i$ is the photometric error, $\delta y_i$ is the spectroscopic 
error  and N is the number of photometric points. 
The constant A 
adjusting the normalization of the two curves can be obtained analytically 
from the condition of minimizing the function with respect to A:
\begin{equation}
   A = {{S_{xy} + (S_{xy}^2 + 4 S_{x3y} S_{xy3})^{1/2}} \over {2 S_{xy3}}}, 
\end{equation}
with the coefficients given by
\begin{eqnarray}
S_{xy} & = &\sum\limits_{i=1}^N (x_i^2\delta y_i^2 - y_i^2 \delta x_i^2) \nonumber \\
S_{xy3} & = &\sum\limits_{i=1}^N x_i y_i \delta y_i^2 \nonumber \\
S_{x3y} & = & \sum\limits_{i=1}^N x_i y_i \delta x_i^2. 
\end{eqnarray}

The method works properly only in the case when the searched time delays are smaller than a half of 
the campaign duration. 
For longer delays the method picks up the longest values of the time delay since this leaves 
only relatively short data sequences, where the variability is low due to the red noise character 
of the curves, and then the two curves can be fitted very well independently of their actual shape. 

\subsubsection{ZDCF method}

We also analyzed the delay between the simulated photometric and spectroscopic
curves using the discrete correlation function (DCF). The idea is rather
simple: all possible pairs are formed from two observation sets and then they
are binned according to their time lag. The hitch is in the details of
statistical evaluation of the bin mean and variance. Here we rely on the ZDCF
flavor of the
method (Alexander 1997), where data variance is estimated directly from
internal scatter within the bin and not from external error estimates. 
This method was also used by Kaspi et al. (2007) in their analysis of the HET
reverberation results.

The original f77 code by Dr. Alexander was reimplemented in f90 and f2py to 
form part of the python time series package (Schwarzenberg-Czerny, 2012). 
Similarly, our fortran simulation code was called
via python wrapper. In doing so we took care to preserve continuity of our 
random series between separate fortran calls.

For each of 1000 simulations of light curves calculated for a fixed parameter
set we analyzed the DCF obtained from the ZDCF algorithm. To
decrease scatter of the DCF it was smoothed with the 3-point running median
filter. For each smooth DCF we determined its
maximum for positive lags $\lambda_m$ and attempted to fit it with a Gaussian
centroid function $f(\lambda)$ with 4 parameters $p_i, i=1,\cdots,4$, i.e.
$f(\lambda)=p_0+p_1\exp\left[-\frac{(\lambda-p_4)^2}{2p_3^2}\right].$ The fit
performed with the python \verb|scipy.optimize.curve_fit| succeeded on 98 \%
occasions.  So although the centroid could have been calculated for the 
majority of the cases, we decided to use the maximum since it works always
and in general did not bring problems.

The results of simulations yielded histograms of the maximum lag $\lambda_m$
and the centroid lag $\lambda_c\equiv p_4$ statistics (Fig.~\ref{fig:histo_ICCF}). From them we
conclude that the maximum of DCF statistics $\lambda_m$ yields the more precise
and robust estimate of the true delay. As far as the limited scope of our
simulations permits to establish, the scatter of the $\lambda_m$ statistics
around its peak is less than that of $\lambda_c$. Additionally, within
precision of our simulations we were not able to find any bias of the estimated
delay (lag of the histogram peak).

\subsection{Dependence on the source variability}

For a fixed method, the accuracy of the measurement depends on the adopted level 
of the variability, as seen from comparison of Table~\ref{table:setup2} and ~\ref{table:setup1}.
Our dimensionless variability amplitude does not translate directly to $F_{var}$ since we
exponentiated the curves to assure the proper statistical behavior, including
the reproduction of the linear dependence of the rms-flux. The actual $F_{var}$ of
our lightcurves is thus slightly higher, about 15\% in 5 yr timescale for adopted 0.1 level.
Since quasars monitored by HET (Kaspi et al. 2007) showed  $F_{var}$ within
the range from 0.03 to 0.124, we also calculated examples of the setup A for amplitudes
of 0.05 and 0.03, using the $\chi^2$ method. The recovered delays were  $655 \pm 204 $ and $635 \pm 232$, 
correspondingly. If the variability level is indeed that low, the accuracy of the measurement should be 
higher to asses the requested accuracy of the delay measurement. Also still better, 
more advance method of the delay measurement, as those based on damped random walk process 
(see Zu et al. 2011, Grier et al. 2012ab) can also help to 
reduce the expected dispersion in a single measurement.  

\subsection{Dependence on the BLR transfer function}

In the previous section we simply parameterized the reprocessing by the BLR 
by a shift and convolution with a Gaussian of a fixed width of 10 \% of the 
delay. Zu et al. (2011) used a top-hat model of a
transfer function. However, the BLR is quite complex and the actual transfer function is
more complicated. Older papers already pointed towards considerable 
difficulties in solving this inversion problem (e.g. Krolik \& Done 1995) 
but already indicated Keplerian component of the motion with considerable 
turbulence and net inflow (analysis of the CIV in NGC 5548, Done  \& Krolik 
1996). In high quality recent data the true BLR complexity shows up pointing 
towards inflow, outflow and the presence of the virialized gas in different 
proportions in different objects (e.g. Denney et al. 2009, Bentz et al. 2008, 
Bentz et al. 2009b, Kollatschny 2003, Grier et al. 2012ab).

Recent high quality reverberation results for nearby Seyfert galaxies allowed 
to reconstruct  the transfer function for a spectral range including 
H$\beta$ line (Bentz et al. 2010, Grier et al. 2013). For most of the objects, at the centroid of H$\beta$, the transfer fuction is roughly Gaussian, but the width varies from $\sim$ 20 \% to 60\% 
(for Mrk 1501).  Therefore, for our representative case A, amplitude 0.1, and $\chi^2$ method 
we calculated the time delay accuracy for increasing width of the Gaussian response. The results
are given in Table~\ref{table:sigma}. The accuracy of the delay measurement slowly decreases with
broadening of the BLR transfer function but the effect is not strong. The red noise character of the 
variability leads to non-monotonic dependence even for 1000 statistical realizations. If the lightcurve happens to have well defined
strong minimum or maximum the measurement is much more accurate than when there is no single 
strong feature. This is also expected to be the case in real quasar lightcurves.

\begin{table*}
\caption{The accuracy of the delay recovery for setup A, variability amplitude 0.1, assumed delay 660 days, as a function of the Gaussian width of the BLR responce}   
\label{table:sigma}      
\centering                          
\begin{tabular}{l r r r r r r r r r}        
\hline\hline      
Model                                &    A & A & A & A & A  \\
Parameter               &    \\    
\hline                        
Width of the BLR response            & 15 \%         & 20\%              & 30 \%    & 40 \%    & 50 \%    &  60 \%  \\
\hline                        
$\chi^2$ \\
Obtained delay                        & $678\pm148$ & $ 688\pm 135$  & $ 690\pm 157$  & $680 \pm 181$  &  $676 \pm 194$ &  $ 689\pm 157$ \\
Spread                                & $122$        & $121$           & $149$          & $182$           &  $204$         & &  $149$      \\
\hline                        
\hline                                   
\end{tabular}
\end{table*}
%

Other authors, like Bottorff et al. (1997), Pancoast et al. (2011) and Goad et al. (2012) used a
physically justified model of the BLR structure which allows to create a library of velocity-resolved or unresolved transfer functions for comparison with the data. Velocity-unresolved transfer function has actually a Gaussian shape for a face-on disk but it starts at zero lag for a ring (Pancoast et al. 2011). Goad et al. (2012) predict very shallow transfer function with a peak at intermediate delays for typical parameters.  

We incorporated exemplary transfer function from Goad et al. (2012), their Fig. B2, inclination $30^{\circ}$, isotropic continuum or disk-like irradiation. The peak of the transfer function in Goad et al. was at 15 days so we simply rescaled the transfer function by a factor 44 to have the peak at 660 days, as in previous simulations. For isotropic continuum, with observational setup A, we recover the delay, although with a considerable error ($678 \pm238 $), and the error is not much smaller for disk-like irradiation ($680 \pm 230$). 

The models above are parametric, without the physics underlaying the shape of the bowl. Therefore, considerable theoretical effort in this directions is clearly needed, as the high quality data for tests start to become available.

\section{Discussion}

We consider in detail the prospects of the determining the dark energy distribution from 
the reverberation studies of high redshift quasars, as proposed by Watson et al. (2011). 
Our motivation comes from the observational campaign with this aim, done with the SALT telescope.
However, measuring the delay between  UV lines and the continuum in high redshift 
quasars is a telescope time consuming project and so a careful analysis performed 
beforehand can help to optimize it.

Our Monte Carlo simulations of the planned monitoring program of high redshift 
AGN show that the success of the monitoring crucially depends on three factors:
\begin{itemize}
\item {the choice of the mathematical method to determine the delay}
\item {the measurement accuracy}
\item {the choice of the emission line}
\end{itemize}
but it is also affected to some extend by other factors like the campaign duration and 
the  distribution of the times of spectroscopic and photometric measurements. These in fact 
are related to some extent to the choice of the quasars for monitoring.
The simulated observational setup mimicked well what we can have for a single quasar 
with a large telescope, including sparse unevenly distributed spectroscopic observations 
with gaps for the invisibility of the source.

The choice of the method used to determine the delay does not have to be made in advance, 
but the estimate of the accuracy achievable in the whole project depends directly on this decision.

We have checked three methods: ICCF (Interpolated Cross Corelation Function), 
ZDCF (Z-transformed Discreet Cross Correlation Function) and $\chi^2$ fitting. 
The first method almost always gave the largest error. The ZDCF was much better, although 
occasionally it also failed due to development of the strong low values tail of the time 
delay histogram. The results were somewhat better, if a Gaussian centroid fit was used 
instead of a maximum of the histogram, but in a few per cent of cases the centroid method failed. 
The $\chi^2$ method was usually the most stable one, although it showed a tail at large delays. 
Therefore, it seems that the best strategy for a real observation is to use both ZDCF and $\chi^2$, 
and compare the results. Still more advanced method based on modeling stochastic processes was introduced by Zu et al. (2011) and its application to a number of sources also show that it gives much lower errors than CCF (Grier et al. 2012ab). We did not use this method with our simulated data but certainly it should be included in the package when analyzing expected observational data.

The accuracy of the photometric and spectroscopic measurements is particularly important 
if the variability of the quasar is low (observed $F_{var}$ less than 0.2 in a five year campaign). 
If the campaign last much longer and/or variability is much stronger then the measurement errors 
are less important, since the red noise variability in the lightcurves will dominate the measurement scatter.

The choice of the emission line for monitoring is absolutely crucial. Our simulated results 
given in Tables~\ref{table:setup2} and ~\ref{table:setup1} were obtained assuming that 
there is no strong intrinsic variability in the line itself, and the observed variability is 
due to reprocessing of the variable continuum, broadened with a Gaussian response of the BLR. 
Indeed, it seems that Low Ionization Lines do not show intrinsic variability, and this was 
the source of the success of the H${\beta}$ monitoring. We plan to use another LIL -- Mg II 2978 \AA~ 
which is also likely just to reflect the nuclear variability. However, if one wants to use 
the CIV line, which is actually necessary for monitoring quasars at redshifts $z>2$, 
a problem appears. CIV lines belongs to the HIL class, and so comes from a different part of the BLR. 
As evidenced by the results of the HET campaign (Kaspi et al.\ 2007), 
in most cases $F_{var}$ of the CIV line was much higher than $F_{var}$ of the continuum, 
implying a strong intrinsic variability of the line,
and only one tentative measurement was obtained. This is further discussed in the next sections.

   \begin{figure}
   \centering
   \includegraphics[width=\hsize]{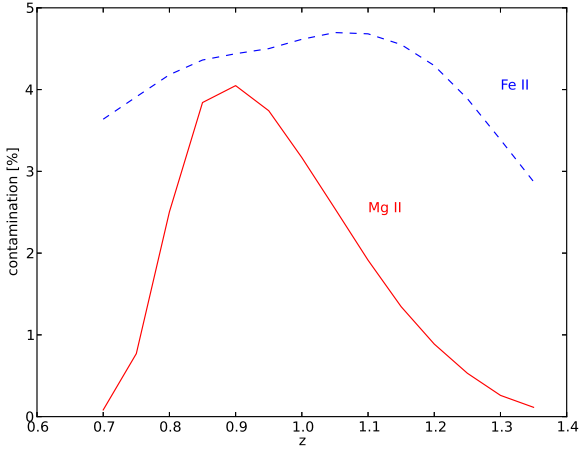}
   \caption{The contamination of the V band photometry by Mg II 2800 \AA~ and Fe II pseudo-continuum as a function
of the quasar redshift.}
              \label{fig:kontaminacja}%
    \end{figure}
%
%

On the other hand, monitoring in Mg II also poses additional problems. The continuum variations are
followed using the broad band V photometry. At some redshift range this leads to contamination of the
continuum measurement by the line itself, and more importantly, by the underlying Fe II pseudo-continuum.
We modeled the problem representing Mg II as a line of fixed EW of $20 $\AA~ and for fixed kinematic width
FWHM of 2500 km s$^{-1}$. To account for Fe II contamination in V photometric filter we used empirical
templates of iron emission bands. Vestergaard \& Wilkes (2001) construct
their templates by identifying Fe II emission in I Zw 1 AGN in the range
1100-3100 A with the most significant emission covering the range from 2000
A to 2900 \AA. To extend iron coverage to 3500~\AA~ we incorporated Tsuzuki et
al. (2006) template which is based on emission fitting in nearby quasars.
Those two templates show very good agreement in their overlapping range.
This allowed us to effectively trace iron contribution to V band in the
redshift range interesting for us, 0.8-1.4. While the Mg II contamination is clearly unimportant, Fe II contributes up to
12 \% of the flux to the V band for a source at the redshift of 0.9 but it drops down below 3\% at $z > 1.3$ 
(see Fig.~\ref{fig:kontaminacja}).

\subsection{MgII vs. CIV line properties and prospects of monitoring}
\label{sect:CIV}

Mg II line properties correlate well with the properties of the Balmer lines in large quasar 
samples (Salviander et al. 2007; McGill et al. 2008; Shen et al. 2008; Wang et al. 2009;  
Shen \& Liu 2012). The line was successfully used for black 
hole mass determination
from a single epoch method  by a number of authors (McLure \& Jarvis 2002;
Vestergaard \& Osmer 2009, Shen et al. 2011).

Woo (2008) showed that for Mg II line the variability level is in general comparable to the 
continuum variability. However, one out of 5 studies quasars (the object CTS300)
actually showed much higher
variability in Mg II line (rms of 17 \% ) than in 3000 \AA ~ continuum (rms of 7\%).
The reason for this variability is unclear since otherwise the source have similar parameters
to the four other objects. 

CIV line properties poorly correlate with Balmer lines in a given object (Baskin \& Laor 2005; Netzer et al. 2007
Shen \& Liu 2012) although some initial studies did not indicate considerable problems 
(Vestergaard 2002, Vestergaard \& Peterson 2006). We discuss this issue in detail below.

\subsection{The problem of CIV monitoring}

   \begin{figure}
   \centering
   \includegraphics[width=\hsize]{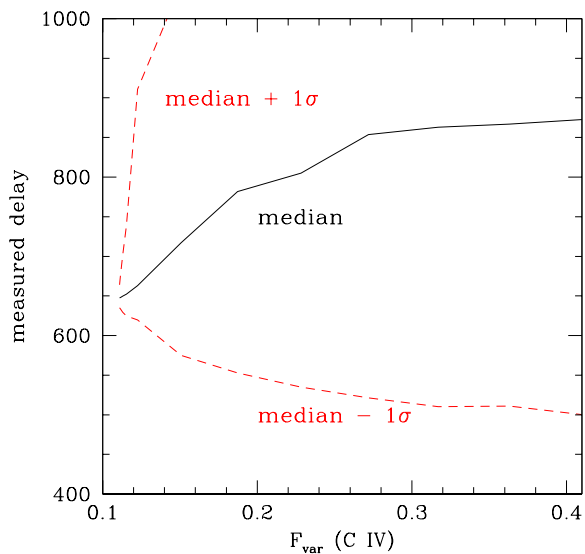}
   \caption{The predicted time delays in 1000 realizations of the model for the use of $\chi^2$ method, with 
the observational setup for HET monitoring and assumed 660 day delay inserted in simulations. There is a 
considerable pile in the last two bins of the delay histogram at the longest time delay which increases with 
increasing intrinsic variability from 5 \% to 26\%.}
              \label{fig:HET}%
    \end{figure}
%
%

In the reverberation project for high redshift quasars with the HET telescope 
and CIV line
(Kaspi et al.\ 2007) the obtained results were inconclusive, with only one 
measurement of a tentative delay of 595 days in S5 0836+71. 
Here we reproduce the key properties of the monitoring: 
(i) objects were observable for six months only; typically (not always) 3 measurements were done, 
separated by 2.5 months, 
(ii) for S5 0836+71 the continuum variability amplitude was $F_{\rm var} = 0.124$, while
the  mean uncertainty for continuum measurements was 2\%, and 10\% for the CIV line. 

We calculate the predicted delay as a function of the CIV variability level, 
considering that the line shows also intrinsic variability in addition to
the response to variable continuum.

We model the intrinsic line variability by formally increasing its measurement error, 
and we performed computations for such setup using the $\chi^2$ method, which usually give 
the best results. The variability level of the continuum is only 
slightly lower than in our model A, low variability case (see Table~\ref{table:setup2}),
 but the spectroscopic and photometric covering is much lower. For the  $F_{var}({\rm CIV}) = 0.233$ 
the recovered delay is $ 805 \pm 265$. One sigma error is large (see Fig.~\ref{fig:HET}), 
and the upper limit is not well constrained, with 23.7 \% of the delays stored in the 
last two histogram bins. The main problem here is not the spectroscopic coverage: without 
the large intrinsic variability we would recover the delay $650 \pm 36$ days for the same 
observational pattern.

This means that CIV monitoring in these sources would require unrealistically long observations for the intrinsic 
red noise variability of the continuum to dominate over the intrinsic variations of CIV. 
Therefore, a better way would be still to use the Mg II line and do the line observations in 
the near-IR band  in the case of sources with considerable intrinsic variability of CIV.

However, in the case of low redshift Seyfert galaxies NGC 5548 (Clavel et al. 1991) , NGC 3783 (Reichert et al. 1994)  the delay of CIV was actually better constrained that the delay of Mg II, and in these two sources the level of
variations in CIV was not higher than in the continuum.  It might be connected to the recent result of Kruczek et al. (2011) that quasars showing hard X-ray spectra, and thus relatively similar to Seyfert 1 galaxies  do not show strong signatures  of outflows while quasars having soft X-ray spectra, more similar to Narrow Line Seyfert 1 galaxies, do display large CIV shifts.  Two classes of quasars  - Population A and Population B - were also discussed  by Sulentic and collaborators in a number of papers (see e.g. Sullentic et al. 2007). The behaviour is likely controlled by the source Eddington ratio, as argued by Wang et al. 2011 at the basis of their comparative study of CIV and Mg II lines in a sample of AGN selected from SDSS. Indeed, the sources selected by the HET team in general show rather narrow Mg II lines. Two sources, S4 0636+68 and S5 0836+71 were studied in the IR, and the FWHM of Mg II was determined to be 2850 km s$^{-1}$ and 3393  km s$^{-1}$, correspondingly  (Iwamuro et al. 2002). For the remaining four objects studied spectroscopically by Kaspi et al. (2007) we roughly estimated the FWHM using SDSS data (Adelman-McCarthy et al. 2007) (SBS 1116+603 $\sim$ 4000  km s$^{-1}$,  SBS 1233+594 $\sim$ 4500 km s$^{-1}$, SBS 1425+606 $\sim$ 7520 km s$^{-1}$, HS 1700+6416 $\sim$ 5300 km s$^{-1}$, and the object with the very broad line indeed showed the comparable variability in the line and the continuum. In addition, the first three objects are rather radio active, with the flux at $10^{10}$ Hz at a similar level than in radio loud composite of Elvis et al. (1994) in comparison with the optical band, and for the remaining three objects there was no radio data in NED. This means that better understanding of the coupling between CIV variability and the remaining properties in clearly needed, and with proper selection the monitoring of distant quasars with CIV can be successful.


\subsection{Consequences for cosmology}

Our simulations indicate that the reverberation measurements in Mg II line for quasars at $z > 1$
can be succesful. This is an interesting project in itself, but the basic goal is to apply these results
to cosmology. This requires not only the determination of the time delay between the Mg II line and 
continuum for each quasar, but one essential further step. Having the delay, we determine the absolute
quasar monochromatic luminosity of a given quasar, as outlined below. At this point quasars can be used
in cosmology in the same way as SN Ia. The essential steps are the following.

For every quasar we plan to determine the observed monochromatic 
luminosity directly from photometry, the size of the BLR, $R_{BLR}$, directly from the time delay,
and the intrinsic (absolute) monochromatic luminosity in our method
is now obtained from the relation
\begin{equation}
\log R_{BLR}[MgII]  = 1.495 + 0.5 \log L_{44,2897},
\label{eq:RBLR}
\end{equation}
in the quasar rest frame. Here we recalculated the relation from Czerny 
\& Hryniewicz (2011) to the Mg II line wavelength. This relation is based on 
assumption of the disk effective temperature 
 $T_{eff} = 1000 $ K at the onset of the BLR, due to the dust formation in the disk atmosphere.
If the dust properties are indeed
independent of redshift,  there is no error in the constants in this formula.

The size of the BLR is measured from the line--continuum time delay, and the accuracy is given in 
Tables~\ref{table:setup2} and \ref{table:setup1}. Adopting a realistic setup (model A, $\chi^2$ method)
we thus gave 5\% accuracy in photometry and 17\% accuracy in time delay, which translates to
the 17\% error in the luminosity distance. If we assume that much better accuracy can be 
actually achieved (model B, $\chi^2$ method) the final error reduces to 3\%. 

It is more convenient to express it using the distance modulus (in magnitudes)  defined as
\begin{equation}
\mu = m-M = 5 \log d_L(z) + 25\;,
\end{equation}
where $d_L(z)$ is the luminosity distance in megaparsecs and $m$ and $M$ are respectively 
the apparent and absolute luminosity of the source.
The accuracy achievable in our method for a single source translates to 0.34 mag in 
the first case and 0.06 mag in the second case, for a single object. 
By monitoring more objects we can increase the accuracy.

Our reverberation project can be thus applied to determine the dark energy distribution much in the same way as SN Ia, the key question is only whether the accuracy of the measurement of the absolute luminosity, or the distance modulus, is high enough to tell anything about the cosmological models. 

\begin{figure}
   \centering
   \includegraphics[angle=270,width=0.5\textwidth]{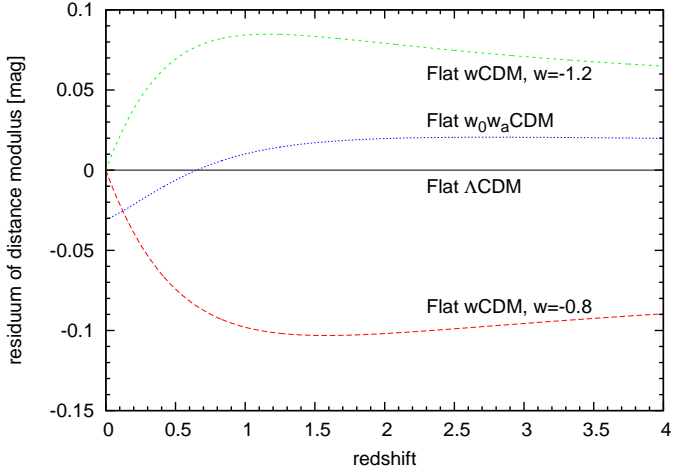}
   \caption{Residua of distance moduli $\mu_{mod}-\mu_{fid}$ (in magnitudes) with respect to the fiducial $\Lambda$CDM model for two wCDM models with constant
equation of state and one $w_0w_a$CDM model with variable $w(z)$.}
              \label{fig:cosmo}%
    \end{figure}

In order to see whether  this accuracy will likely bring interesting results we
compare it with several cosmological models. For a
given cosmological model one can compute the distance modulus as a function of redshift and 
compare it with observations. In Figure \ref{fig:cosmo}
we plot residua of distance moduli with respect to the fiducial $\Lambda$CDM model, 
$\mu_{\rm mod}-\mu_{\rm fid}$, assuming null spatial curvature
(generalization to non-flat models is straightforward). We show two wCDM models with 
a constant equation of state of dark energy $w=-0.8$ and
$w=-1.2$, and one with variable $w(z)=w_0+w_a z/(1+z)$ (the '$w_0w_a$CDM' model). The 
fiducial cosmological parameters used were the central values obtained by Mehta et al.\ (2012): 
$\Omega_{\rm m}=0.28$ and $H_0=70$ km s$^{-1}$Mpc$^{-1}$ in case of the $\Lambda$CDM and wCDM models, 
and $\Omega_{\rm m}=0.272$, $H_0=71$km s$^{-1}$Mpc$^{-1}$, $w_0=-1.02$, $w_a=-0.26$ for the $w_0w_a$CDM model.

Since the expected differences between different cosmological models are of the order of 0.1 mag, 
it is possible to test the cosmological models with quasars at $z \sim 1$, 
if the measurement accuracy and/or the number of sources is high.

%
%

\subsection{Limitations of the method, sources of systematic errors and future method developments}

Quasars are now being observed up to redshifts above 7 (ULAS J1120+0641, $z = 7.085$, Mortlock et al. 2011). The monitoring in the optical band based on Mg II line is applicable only to $0.5 <z < 1.5 $ quasars. For higher redshifts, lines like CIV and Ly$\alpha$ must then be used to trace BLR. Our current method using Mg II is based on applicability of Eq.\ref{eq:RBLR}, i.e. on the assumption that lines form close to the atmosphere of a Keplerian disk, and the line emitting clouds should not show systematic departure from the Keplerian motion as well, apart from a turbulent motion up and down due to subsequent condensation and evaporation of the dust (Czerny \& Hryniewicz 2011). Low Ionization Lines like H$\beta$ and Mg II satisfy this condition approximately although some evidence for inflow is seen even in the Balmer lines (Grier et al. 2013). In the case of CIV the evidence of outflow is well established, and it is strong in some quasars but moderate in others (see Kruczek et al. 2011 for recent determinations of CIV blueshift and its relation with the overall quasar spectrum). This motion has to be taken into account when relating the black hole mass, monochromatic luminosity and the position of the specific line, so Eq.\ref{eq:RBLR} has to be replaced by more complex theory of the BLR formation. In the case of a specific object, also the viewing angle is an issue although in statistical approach, using many objects, this may average to some extend. Systematic errors of the method are now difficult to specify but further studies of the transfer function in the nearby sources and corresponding development of the BLR modeling are absolutely necessary. 

The second assumption underlying  Eq.\ref{eq:RBLR} is the assumption that the dust properties are universal in all quasars. Dust provides the driving force behind the rise and fall of the clouds and explains large covering factor of the BLR, and if dust composition is a function of redshift the dust evaporation temperature would also depend on the quasar redshift. Specifically, since graphite grains have higher evaporation temperature and silicates and amorphous grains have lower evaporation temperature, the mean effect depends on the chemical composition of the disk atmosphere. fortunately, the knowledge of nearby sources can likely apply to high redshift quasars due to the small metallicity gradient with redshift in quasar population, and all high Eddington luminosity sources have somewhat super-solar metallicity, both locally and for high redshift sources (e.g. Simon \& Hamann 2010).

\subsection{Comparison of quasar method with other methods}

The current dark energy constraints already come from various research lines. All methods can be divided into
standard ruler methods and standard candle methods. Standard ruler methods relay on determination of a  true or characteristic size and measurements of the apparent size, and most current methods belong to this class ( microwave background modeling, all large scale structure methods including BAO, power spectra and void studies, weak lensing, galaxy cluster methods). Our method, based on determination of the quasar intrinsic luminosity, clearly belongs to the second family, together with the supernovae Ia and gamma-ray burst methods.  If considered in the $\Omega_{\Lambda} $ vs. $\Omega_{m}$ space these methods usually give considerably elongated contour errors, roughly perpendicular to each other. This is the reason why the combination of the two methods is usually used to give strong constraints on the cosmological models. Our method has the same in-built degeneracy as the Supernovae Ia, and it would also need to be combined with complementary constraints (e.g. new results from Planck mission) to break it. 

The advantage of the method is that it can easily cover somewhat higher redshift range than supernovae (the groud-based observations of supernovae Ia do not extend beyond $z = 1$, Regnault  et al. 2009;  the HST observations are needed to  bring objects  $0.623 < z < 1.415 $, Suzuki et al. 2012). Methods based on gamma-ray bursts can go much further but they still need much more study to achieve the proper understanding. Quasars are bright and numerous. On the other hand, the reverberation studies are indeed time consuming, and the delays increase with redshift unless we go to fainter objects containing less massive black holes than the extreme bright sources.  Thus single-object observations like long-slit spectroscopy with SALT cannot bring many measurements. However, multi-object spectroscopic studies can bring the data suitable for application of our method. Current instruments like  LAMOST may in principle perform such studies (see e.g. Wu et al. 2010) if working at full capacity. EUCLID among the planned missions with lifetime of 6 years would be also perfect for that purpose although  the mission is rather aimed at covering large area of the sky without many repeated observations of the same area.


\begin{acknowledgements}
      Part of this work was supported by the Polish grant Nr. 719/N-SALT/2010/0, and by the grants 
2011/03/B/ST9/03281 and DEC-2011/03/B/ST9/03459 from the National Science Center. BCZ and KRHR acknowledge the support by the Foundation for Polish Science through Master/Mistrz program. The financial assistance of the South African National Research Foundation (NRF) towards this research is hereby acknowledged (MB). ASC was funded by grant NCN 2011/01/M/ST9/05914.  This publication makes use of data products from the Sloan Digital Sky Survey. Funding for the SDSS and SDSS-II has been provided by the Alfred P.
Sloan Foundation, the Participating Institutions, the National Science
Foundation, the U.S. Department of Energy, the National Aeronautics
and Space Administration, the Japanese Monbukagakusho, the Max Planck
Society, and the Higher Education Funding Council for England. The
SDSS Web Site is http://www.sdss.org/.
This research has also made use of the NASA/IPAC Extragalactic Database (NED) which is operated by the Jet Propulsion Laboratory, California
Institute of Technology, under contract with the National Aeronautics and Space Administration.

\end{acknowledgements}


\begin{thebibliography}{}

\bibitem[]{} Ade, P.A.R. et al., 2012, arXiv:1303.5076
\bibitem[]{} Adelman-McCarthy J.~K. et al. 2007, ApJS, 172, 634
\bibitem[]{} Alexander, T., 1997, in ``Is AGN Variability Correlated with Other AGN Properties?
ZDCF Analysis of Small Samples of Sparse Light Curves
in Astronomical Time Series'',
Eds. D. Maoz, A. Sternberg, and E.M. Leibowitz, Astrophysics and Space Science Library, 218, 163
\bibitem[]{} Amendola, L., Tsuikawa, S., 2010, Dark energy, Cambridge University Press
\bibitem[]{} Barbieri, C., Cappellaro, E., Turatto, M., Romano, G., Szuszkiewicz, E., 1988,
A\&AS, 76, 477
\bibitem[]{} Barth A.J et al., 2011, ApJ, 732, 121
\bibitem[]{} Baskin, A., \& Laor, A. 2005, MNRAS, 356, 1029
\bibitem[]{} Bauer, A. et al., 2009, ApJ, 696, 1241
\bibitem[]{} Benson, B.A. et al., 2013, ApJ, 763, 147
\bibitem[]{} Bentz, M. C., et al. 2008, ApJ, 689, L21 
\bibitem[]{} Bentz, M. et al. 2009a, ApJ, 697, 160
\bibitem[]{} Bentz, M. et al., 2009b, ApJ, 705, 199 
\bibitem[]{} Bentz, M. et al., 2010, ApJL, 720, L46
\bibitem[]{} Bian, W.,  Yang, Z., 2011, Journal of Astrophysics and Astronomy,  32, 253
\bibitem[]{} Biesiada, M., Piorkowska, A., 2009, MNRAS, 396, 946
\bibitem[]{} Bladford, R.D.,  McKee, C.F., 1982, ApJ, 255, 419
\bibitem[]{} Bottorff, M.C., Korista, K.T., Shlosman, I., Blandford, R.D., 1997, ApJ, 479, 200
\bibitem[]{} Chelouche, D., Daniel, E., Kaspi, S., 2012, ApJ, 750, L43
\bibitem[]{} Chevallier, M., Polarski, D., 2001, IJMPD, 10, 213
\bibitem[]{} Clavel, J., et al., 1991, ApJ, 366, 64
\bibitem[]{} Cristiani, S. et al. 1996,  A\&A, 306, 395
\bibitem[]{} Czerny, B., Hryniewicz, K., 2011, A\&A, 525, L8
\bibitem[]{} Czerny B., Nikolajuk M., Piasecki M., Kuraszkiewicz J., 2001, MNRAS,
325, 865
\bibitem[]{} Czerny, B. et al., 2003, MNRAS, 342, 1222
\bibitem[]{} Czerny, B., Schwarzenberg-Czerny, A., Loska, Z., 1999, MNRAS, 303, 148 
\bibitem[]{} Dainotti, M. G., Ostrowski, M., Willingale, R.,  2011, MNRAS, 418, 2202
\bibitem[]{} Denney, K.D. et al. 2009, ApJ, 702, 1353
\bibitem[]{} Denney, K.D. et al., 2010, ApJ, 721, 715
\bibitem[]{} De Rosa, G.  et al.,  2011, ApJ, 739, 56
\bibitem[]{} Dietrich, M., Hamann, F., Appenzeller, I., Vestergaard, M., 2003a, ApJ, 596, 817
\bibitem[]{} Dietrich, M. et al. 2003b, ApJ, 589, 722
\bibitem[]{} Dietrich, M. et al., 2012, ApJ, 757, 53 
\bibitem[]{} Done, C., Krolik, J.H., 1996, ApJ, 463, 144
\bibitem[]{} Edelson, R. A. et al., 1996, ApJ, 470, 364
\bibitem[]{} Eisenstein, D.J. et al., 2005, ApJ, 633, 560
\bibitem[]{} Elvis, M. et al., 1994, ApJS, 95, 1
\bibitem[]{} Frieman, J. A., Turner, M. S., Huterer, D., 2008, ARA\&A, 46, 385
\bibitem[]{} Gall, C., Andersen, A. C., Hjorth, J.,  2011, A\&A, 528, A14
\bibitem[]{} Gaskell, , C.M., Peterson, B.M., 1987, ApJS, 65, 1
\bibitem[]{} Goad M. R., Korista K. T., Ruff A. J., 2012, MNRAS, 426, 3086
\bibitem[]{} Grier, C. et al. 2012a, ApJ, 744, L4
\bibitem[]{} Grier, C. et al. 2012b, ApJ, 755, 60 
\bibitem[]{} Grier, C. et al. 2013, ApJ, 764, 47
\bibitem[]{} Guzzo, L. et al., 2008, Natur, 451. 541
\bibitem[]{} Hawkins, M. R. S. 1983, MNRAS, 202, 571
\bibitem[]{} Hayashida K., Miyamoto S., Kitamoto S., Negoro H., Inoue H., 1998, ApJ,
500, 642
\bibitem[]{} Hewett, P.C., Foltz, C.B., Chaffee, F.H. 1995, AJ, 109, 1498
\bibitem[]{} Hinshaw, G. et al., 2012, arXiv:1212.5226
\bibitem[]{} Iwamuro, F. et al., 2002, ApJ, 565, 63
\bibitem[]{} Jain, B., Taylor, A., 2003, Phys. Rev. Lett., 91n1302
\bibitem[]{} Kaspi, S. et al. 2000, ApJ, 533, 631
\bibitem[]{} Kaspi, S. et al. 2007, 659, 997
\bibitem[]{} Kofman, L.A.,  Gnedin, N.Y., Bahcall, N.A., 1993, ApJ, 413, 1
\bibitem[]{} Kollatschny, W. 2003, A\&A, 407, 461
\bibitem[]{} Kong M.-Z., Wu X.-B., Wang R., Han J.-L., 2006, Chin. J. Astron. Astrophys., 6, 396
\bibitem[]{} Krolik, J.H., Done, C., 1995, ApJ, 440, 166
\bibitem[]{} Kruczek, N. E. et al., 2011, AJ, 142, 130
\bibitem[]{} Li, M., Li, X.-D., Wang, S., Wang, Y., 2011, arXiv:1103.5870v6
\bibitem[]{} Mattarese, S., Colpi, M., Gorini, V., Moschella, U. (Editors), 2011, Dark Matter and Dark Energy, Astrophysics \& Space Science Library, Springer, Netherlands
\bibitem[]{} McGill, K. L., Woo, J., Treu, T., \& Malkan, M. A. 2008, ApJ, 673, 703
\bibitem[]{} McHardy I. M., Koerding E., Knigge C., Uttley P., Fender R. P., 2006, Nat,
444, 730
\bibitem[]{} MacLoad, C.L. et al., 2012, ApJ, 753, 106 
\bibitem[]{} Matthews, T.A., \& Sandage, A.R. 1963, ApJ, 138, 30
\bibitem[]{} McLure, R. J., \& Jarvis, M. J. 2002, MNRAS, 337, 109
\bibitem[]{} Mehta,, K. T., Cuesta, A. J., Xu, X., Eisenstein, D. J., Padmanabhan, N., 2012, arXiv:1202.0092 
\bibitem[]{} Metzroth, K.G., Onken, C.A., Peterson, B.M., 2006, ApJ, 647, 901
\bibitem[]{} Mortlock D. J. et al., 2011, Nat, 474, 616
\bibitem[]{} Mushotzky, R. F., Edelson, R., Baumgartner, W., Gandhi, P., 2011, ApJ, 743, L12
\bibitem[]{} Netzer, H., Lira, P., Trakhtenbrot, B., Shemmer, O., \& Cury, I. 2007, ApJ, 671, 1256
\bibitem[]{} Nikolajuk, M., Czerny, B., Gurynowicz, P., 2009, MNRAS, 394, 2141
\bibitem[]{} O'Brien P. T. et al., 1998,  ApJ, 509, 163
\bibitem[]{} Pancoast, A., Brewer, B.J., Treu, T., 2011, ApJ, 730, 139
\bibitem[]{} Papadakis I. E., 2004, A\&A, 425, 1133
\bibitem[]{} Peebles, P.J.E., 1984, ApJ, 284, 439
\bibitem[]{} Perlmutter, S. et al., 1998, Natur, 391, 51
\bibitem[]{} Peterson, B.M., 1993, PASP, 105, 247
\bibitem[]{} Peterson, B.M. et al., 2002, ApJ, 581, 197
\bibitem[]{} Peterson, B.M. et al. 2004, ApJ, 613, 682
\bibitem[]{} Peterson, B.M., Wandel, A., 1999, ApJ, 521, L95
\bibitem[]{} Ponti, G., et al., 2012, A\&A, 542, A83
\bibitem[]{} Regnault, N., et al., 2009, A\&A, 506, 999
\bibitem[]{} Reichert, G.A. et al., 1994, ApJ, 425, 582
\bibitem[]{} Riess, A.G. et al., 1998, AJ, 116, 1009
\bibitem[]{} Salviander, S., Shields, G. A., Gebhardt, K., \& Bonning, E. W. 2007, ApJ, 662, 131
\bibitem[]{} Sanchez, A.G. et al., 2012, MNRAS, 425, 415
\bibitem[]{} Schutz, B. F., 1986 Nature, 323, 310 
\bibitem[]{} Schwarzenberg-Czerny, A., 2012, in IAU Symposium 285, 81
\bibitem[]{} Sergeev, S.G., Malkov, Yu. F., Chuvaev, K. K., Pronik, V. I., 1994, ASPC, 69, 199
\bibitem[]{} Seto, N., Kawamura, S., and Nakamura, T., 2001, Phys. Rev. Lett. 87 221103
\bibitem[]{} Shen, Y., Greene, J. E., Strauss, M. A., Richards, G. T., \& Schneider, D. P.
2008, ApJ, 680, 169
\bibitem[]{} Shen, Y., Liu, X., 2012, ApJ, 753, 125
\bibitem[]{} Shen, Y., Richards, G. T., Strauss, M. A., et al. 2011, ApJS, 194, 45
\bibitem[]{} Simon, L. E., Hamann, F., 2010, 2010, MNRAS, 407, 1826
\bibitem[]{} Stalin, C.S. et al., 2011, MNRAS, 416, 225
\bibitem[]{} Sulentic, J. et al., 2007, ApJ, 666, 757
\bibitem[]{} Suzuki, N. et al., 2012, ApJ, 746, 85
\bibitem[]{} Terebizh, V.Y., Terebizh, A.V., Biryukov, V.V., 1989, Ap, 31, 460
\bibitem[]{} Timmer, J.,  Konig, M.,  1995, A\&A, 300, 707
\bibitem[]{} Tsuzuki Y., Kawara K., Yoshii Y., Oyabu S., Tanabé T., Matsuoka
Y., 2006, ApJ, 650, 57
\bibitem[]{} Ulrich, M.-H.; Maraschi, L., Urry, C. M., 1997, ARA\&A, 35, 445 
\bibitem[]{} Uttley, P., McHardy, I. M., Vaughan, S., 2005, MNRAS, 359, 345
\bibitem[]{} Vanden Berk, D.E. et al. 2004, ApJ, 601, 692
\bibitem[]{} Vestergaard, M. 2002, ApJ, 571, 733
\bibitem[]{} Vestergaard, M., \& Osmer, P. S. 2009, ApJ, 699, 800
\bibitem[]{} Vestergaard, M., \& Peterson, B. M. 2006, ApJ, 641, 689
\bibitem[]{} Vestergaard M., Wilkes B.J., 2001, ApJS, 134, 1
\bibitem[]{} Walsh, J.L.,  et al. 2009, ApJS, 185, 156
\bibitem[]{} Wandel. A., Peterson, B.M., Malkan, M.A., 1999, 526, 419
\bibitem[]{} Wang, J.G. et al., 2009, ApJ, 707, 1334
\bibitem[]{} Wang, T.-G., Wamsteker, W.,  Cheng, F.-Z., 1997, in ASP Conference Series, 113, 155 
\bibitem[]{} Wang, H., Wang, T., Zhou, H., et al., 2011, ApJ, 738, 85
\bibitem[]{} Watson, D. et al. 2011, ApJ, 740, L49
\bibitem[]{} Woo, J.-H., 2008, AJ, 135, 1849
\bibitem[]{} Wu, Xue-Bing et al. 2010,  RAA, 10, 745
\bibitem[]{} Zhou, X.-L. Zhang, S.-N., Wang, D.-X., Zhu, L., 2010, ApJ, 710, 16
\bibitem[]{} Zu, Y., Kochanek, C.S., Peterson, B.M., 2011, ApJ, 735, 80

\end{thebibliography}
\end{document}